\documentclass[aip,pop,twocolumn,floatfix,amsmath,amssymb, showpacs,tightenlines]{revtex4}
\usepackage{mathrsfs}
\usepackage{bbm}
\usepackage{amssymb}
\usepackage{graphicx,color}
\usepackage{epsfig,graphicx,times}
\usepackage{amstext}
\usepackage{amsmath}
\usepackage{latexsym}
\usepackage{bm}
\usepackage{CJK}

\begin{document}

\title{High flux symmetry of the Spherical Hohlraum with Octahedral 6LEHs at a Golden Hohlraum-to-capsule Radius ratio}
\author{Ke Lan  $^1$, Jie Liu  $^{1,2}$, Dongxian Lai  $^1$£¬ Wudi Zheng
  $^1$,
  Xian-Tu He $^{1,2}$}

\address{$^1$Institute of Applied Physics and Computational Mathematics, Beijing, 100088,
China\\
$^2$Center for Applied Physics and Technology, Peking University,
Beijing, 100871, China}

\begin{abstract}
In the present Letter, we investigate  a spherical hohlraums with
octahedral six laser entrance holes (LEHs) for inertial fusion,
which has advantages over the conventional hohlraums of cylindrical
geometry since it contains only one cone at each LEH and the
problems caused by the beam overlap and crossed-beam energy transfer
can be eliminated and the backscattering can be reduced.
 In particular, our study indicates that at a specific hohlraum-to-capsule radius ratio, i.e., the golden ratio, the flux asymmetry on
 capsule can be significantly reduced.
From our study, this golden octahedral hohlraum   has robust high
symmetry, low plasma filling and low backscattering. Though the
golden octahedral hohlraum needs  $30\%$ more laser energy than
traditional cylinder for producing the ignition radiation pulse of
300 eV, it is worth for a robust high symmetry and low
backscattering. The proposed octahedral hohlraum is also flexible
and can be applicable to diverse inertial fusion drive approaches.
As an application, we design an ignition octahedral hohlraum for the
hybrid drive.

\end{abstract}


\maketitle

\emph{Introduction}---The hohlraum is  crucial for the inertial
fusions of  both indirect drive \cite{lindl2004, MTV, Haan} and the
hybrid indirect-direct drive proposed recently (HID) \cite{Fan}. In
the indirect drive approach, the hohlraum is first heated by laser
beams to a few million Kelvin and then the energy flux of the
transferred X-ray radiation compress the deuterium-tritium capsule
at a convergence ratio of 25 to 45, making the nuclear fuel finally
burn in a self-sustained way. In the corresponding hohlraum design,
the hohlraum shape, size and the number of Laser Entrance Hole (LEH)
are optimized to balance tradeoffs among the needs for capsule
symmetry, the acceptable hohlraum plasma filling, the requirements
for energy and power, and the laser plasma interactions. Among many
requirements, the energy coupling and flux symmetry are of most
concerned. A higher energy coupling  will economize the input energy
and increase the fusion energy gain. More importantly, a very
uniform flux from the hohlraum on the shell of capsule is mandatory
because a small drive asymmetry of 1$\%$ \cite{lindl2004} can lead
to the failure of ignition. Actually, the small flux asymmetry will
be magnified during the compression process due to the varied kinds
of instabilities and results in a serious hot-cold fuel mixture that
can dramatically lessen the temperature or density of the hot spot
for ignition.

Various hohlraums with different shapes have been  proposed and
investigated, such as cylinder hohlraum\cite{lindl2004, MTV}, rugby
hohlraum \cite{Caruso, Amendt, Vandenboomgaerde, Casner, Robey,
Phillippe} and elliptical hohlraum \cite{Lan2012}. These hohlraums
are elongated with a length-to-diameter ratio greater than unity and
have cylindrically symmetry with two LEHs on the ends. Among all
above hohlraums, the cylindrical hohlraums are used most often in
inertial fusion studies and are chosen as the ignition hohlraum on
NIF \cite{Haan, Callahan, Kline}, though it breaks the spherical
symmetry and leads to cross coupling between the modes.

Intuitively,  spherical hohlraum has the feature of the most
symmetry compared to other geometric shapes. In the late 1990s,
experiments on hohlraum with 6 LEHs obviously exhibited its
advantage in high uniformity of the radiation flux on the capsules'
surface\cite{Belkov}, while the theoretical investigations of this
kind of hohlraum design are in lack. Soon after that, the first
experiment on hohlraum with 4 LEHs of tetrahedral symmetry was
conducted at OMEGA \cite{Wallace} while the theoretical study showed
that it needs two sets of laser beams in order to minimize the flux
asymmetry by varying  the relative power \cite{Phillion}.

 In this Letter, we investigate the
 spherical hohlraum with octahedral 6 LEHs  for the first time from the theoretical side,
 addressing the most important issue of the  flux symmetry.
 We find   a golden hohlraum-to-capsule radius ratio of 5.14, at which the flux asymmetry can be reduced to about
 0.1$\%$. We call the hohlraum as golden octahedral hohlraum. From our
 study, there is a robust high symmetry inside such a golden
 octahedral hohlraum during the capsule implosion.
In addition, the golden octahedral hohlraum  contains only one cone
at each LEH and  the backscattering can be small without any beam
phasing \cite{Haan}. The golden octahedral hohlraum also
 has low plasma filling, which further benefits for a low
 backscattering.
However, A larger volume of the hohlraum needs a little more laser
energy to drive a golden octahedral hohlraum than to drive a
traditional cylinder for generating  same radiation. Nevertheless,
it is worth to exchange some laser energy for  a robust high
symmetry. The octahedral hohlraum   design can be implemented on the
Omega laser and will be conducted on SG laser facilities in 2014.
 As an application, we  design
 a golden octahedral hohlraum for the hybrid drive  using the
 expended plasma-filling model and  view factor model.

\emph{Spherical Hohlraum with Octahedral 6LEHs}---For convenience,
we consider that octahedral hohlraum has two poles and an equator
though it is round. In the octahedral hohlraum, there are six LEHs,
one at per pole and four along the equator coordinately. In the
hohlraum system, we define $\theta$ as polar angle and  $\phi$ as
azimuthal angle. We use $R_H$ to denote the hohlraum radius, $R_C$
the capsule radius, $R_L$ the LEH radius  and $R_Q$ the quad radius
at LEH. Here, we assume the quad shape at LEH  to be a circle. Each
quad through a LEH is characterized by $\theta _L$ and $\phi _L$,
where $\theta _L$ is the opening angle that the quad makes with the
LEH normal direction and $\phi _L$ is the azimuthal angle about the
normal of the LEH. The relative fluxes of the laser spot, the
hohlraum wall and LEH are denoted as $F_{spot}$,
 $F_{wall}$ and $F_{LEH}$, respectively. Usually, we take
$F_{spot} : F_{wall} : F_{LEH}$ = 2 :1 : 0, unless declaring. Fig. 1
shows the scenography of the octahedral hohlraum  with six LEHs and
laser spot of 48 quads and its pattern  in the $\theta$/$\phi$
plane, by taking $R_H/R_C$=5.1, $R_C$=1.1 mm, $R_L$=1 mm, $R_Q$=0.3
mm and $\theta _L = 55^\circ$.  From our calculation, the flux
asymmetry is about $0.1\%$ on a capsule of 1.1 mm radius.

\begin{figure}[t]
\includegraphics[bb=180 450 430 800,width=4 cm]{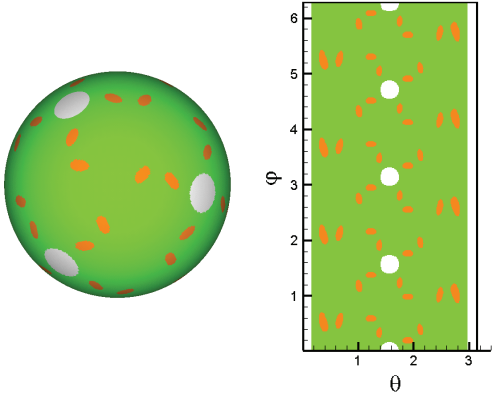}
\caption{(color online) Scenography of octahedral hohlraum  with six
LEHs (white color) and laser spot of 48 quads(red color) on the
left-hand side and its pattern  in the $\theta$/$\phi$ plane on the
right-hand side, by taking $R_H/R_C$=5.1, $R_C$=1.1 mm, $R_L$=1 mm,
$R_Q$=0.3 mm and $\theta _L = 55^\circ$.  } \label{Fig1}
\end{figure}


\emph{Golden ratio}---We firstly  use a simple model to prove that a
golden hohlraum-to-capsule radius ratio exists for an octahedral
hohlraum, at which the flux asymmetry can reach its minimum.
 In this simple model, the LEHs are treated as
negative sources, and the wall and laser spots are treated as a
homogeneous background by neglecting their flux difference.  Only
considering the negative effect of LEH on capsule, we present in
Fig. 2 a schematic of a capsule inside an octahedral hohlraum. The
capsule is concentric with the octahedral hohlraum with their center
at point O. On capsule, there are two kinds of points which see LEH
most different,
 such as points A and B in the figure.
The normal of point A is in the same direction as that of the LEH
centered on point M, while the normal of point B has equal angles
with that of three LEHs, centered on points L, M and N,
respectively. Hence, we can study the flux asymmetry on capsule by
comparing the irradiation on points A and B.

The flux irradiated on a capsule point is mainly decided by the
solid angle of the source opened to that point and the angle of the
connecting line with respect to the normal of  the capsule point. By
denoting the LEH area as  $S$, the solid angle of LEH M seen by
point A is $ d\Omega _A=S/(R_H-R_C)^2 $, and the solid angle of LEH
N seen by point B is $  d\Omega _B=S cos \alpha/l^2$.
 Here, $l$  is the length of line BN,
and  $\alpha$ is the angle of BN with respect to the normal of N. We
use  $\beta$ to denote the angle between the normal of N and the
normal of B, then the angle of BN with respect to the normal of B is
$\alpha + \beta$. Note that both LEH M and LEH L open the equal
solid angle to point B as LEH N. Then the quality of the
illumination on the capsule can be quantified approximately by: $
f=0.5 \times [3 d \Omega _B \times cos (\alpha + \beta) - d \Omega
_A] /d \Omega _A$. From Fig. 2, we have the following geometrical
relationships: $ tg \beta = \sqrt 2 $, $tg \alpha =R_C sin \beta /
(R_H - R_C cos \beta)$ and $ l \times cos \alpha = R_H - R_C cos
\beta $. Then, we obtain:
\begin{eqnarray}\label{Eq10}
f=0.5 \times [3  cos ^3 \alpha \times cos (\alpha + \beta) \times (
\frac {\frac {R_H}{R_C} -1 }{\frac {R_H}{R_C} - cos \beta} )^2 -1 ]
\end{eqnarray}
The variation of $|f|$  as $R_H / R_C$  is presented in Fig. 3. As
shown, $|f|$ reaches its minimum at $R_H / R_C = 5.14$. It predicts
the emergence of the minimum flux asymmetry at the golden
hohlraum-to-capsule radius ratio.

\begin{figure}[t]
\includegraphics[bb=150 560 400 750,width=6 cm]{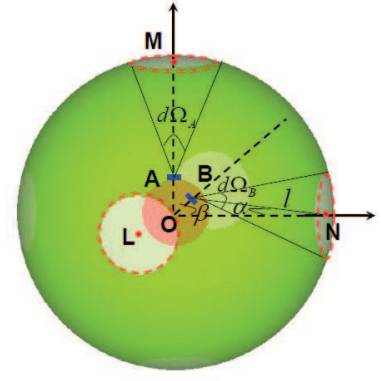}
\caption{(color online) Schematic of a capsule inside an octahedral
hohlraum } \label{Fig3}
\end{figure}

\begin{figure}[t]
\includegraphics[bb=00 340 350 660,width=6 cm]{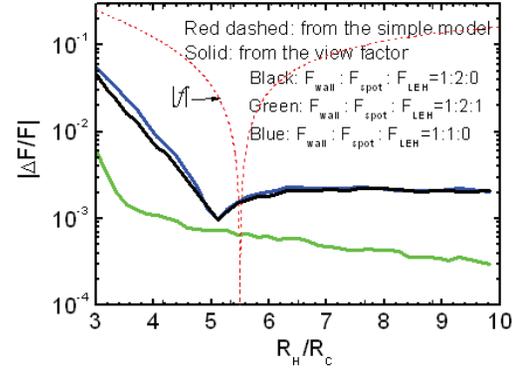}
\caption{(color online) Variations of    $|f|$ as $R_{H}/R_{C}$ from
the simple model (red line) and  $|\Delta F /F|$ from the view
factor model (black solid line). } \label{Fig4}
\end{figure}

\emph{Calculations with view factor model}---To certify the above
theoretical prediction, we further exploit the view factor model to
calculate the radiation flux on the shell of the capsule
numerically. We define ratio $|\Delta F /\langle F \rangle|$, in
which $\Delta F = 0.5 \times (F_{max} - F_{min})$ and $\langle F
\rangle$ is the average value of flux $F$ upon the capsule. The
black solid line shown in Fig. 3 is variation of $|\Delta F /\langle
F \rangle|$ as $R_{H}/R_{C}$ on the capsule shown in Fig.2, which is
inside an octahedral hohlraum with $R_L$=1 mm, $R_Q$=0.5 mm and
$\theta _L = 55^\circ$. As indicated, an asymmetry minimum do exist
for $|\Delta F /\langle F \rangle|$ at $R_H/R_C=5.1$, quite close to
the simple model. Using $F({\bf P})$ to denote the total flux F at
point $\bf P(\theta, \phi)$ on capsule, the asymmetry of flux on
capsule can be expanded as $F({\bf P})=\sum ^{\infty} _{l=0} \sum
^{l} _{m=-l} a_{lm} Y_{lm} (\theta , \phi)$, where $Y_{lm} (\theta ,
\phi)$  are the spherical harmonics and $a_{lm}$  is spherical
harmonic decomposition. We further define $C_{l0} = a_{l0}/a_{00}$
and $C_{lm} = 2a_{lm}/a_{00}$ for $m>0$.  Shown in Fig. 4 is
variations of $C_{lm}$ as $R_{H}/R_{C}$ for the same model in Fig.
3. As shown, $C_{40}$ and $C_{44}$ dominate the capsule flux
asymmetry, except around the golden ratio where the asymmetry is
dominated by $C_{80}$, $C_{84}$ and $C_{88}$  with values much
smaller than $0.1\%$. Notice that $C_{2m}$ is on noise level and can
be thoroughly neglected inside an octahedral hohlraum, quite
different from the case inside a cylindrical geometry. The minimums
of $C_{4m}$ at $R_H/R_C=5$, $C_{6m}$ at around $R_H/R_C=4$ and
$C_{8m}$ at around $R_H/R_C=6$ are due to the asymmetry smoothing
factor on capsule inside a concentric spherical hohlraum\cite{MTV,
lindl2004, Caruso}. According to Ref. [1], the smoothing factor
depends on mode number $l$ but not on the directional mode number
$m$ because the choice of the direction of the polar axis is
arbitrary for spherical symmetry.   Here, it is worth to mention the
2LEH cylindrical and 4LEH spherical hohlraums, in which the
symmetries are dominated respectively  by $Y_{2m}$ and $Y_{3m}$,
while the smoothing factors of $Y_{2m}$ and $Y_{3m}$ are much less
reduced, especially  at $R_H/R_C \leq 5$.

\begin{figure}[b]
\includegraphics[bb=30 10 380 250,width=6 cm]{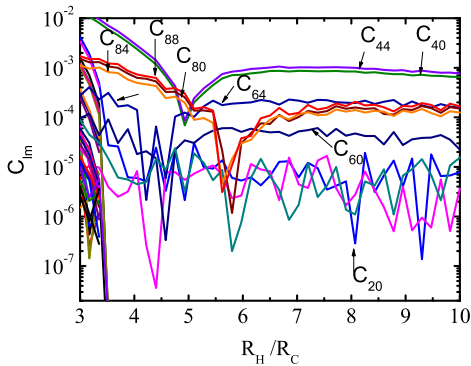}
\caption{(color online) Variations of $C_{lm}$ as $R_{H}/R_{C}$. }
\label{Fig6}
\end{figure}

In order to distinguish the asymmetry contributions from LEH and
laser spot, we calculate spherical hohlraums with only octahedral
6LEHs and only 48 quads, respectively. As shown in Fig. 3, the
asymmetry contributed by the LEHs is significantly larger than that
by the spots, and the asymmetry  is mainly decided by the LEHs.
Obviously, the asymmetry minimum is thoroughly  due to  the six LEHs
of the octahedral hohlraum. That is why $R_H/R_C$ of the minimum
asymmetry from the simple model agrees so well with that from view
factor model. According to our calculations, $R_H/R_C$ of the
minimum asymmetry has small deviation from 5.14 under different
 LEH-to-hohlraum radius ratio and different arrangement of laser
beams. We call $R_H/R_C=5.14$ as the golden ratio of the octahedral
hohlraum.  Notice that $|\Delta F /\langle F \rangle|$ is around
$0.2 \%$ at $R_H/R_C \geq 4.7$, which means that the golden
octahedral hohlraum  has robust symmetry not only at the early stage
of capsule implosion when $R_H/R_C$ becomes a little small due to
wall plasma expansion, spot motion and expansion of the outer layers
of the capsule, but also during the stages of capsule inward
acceleration and ignition when $R_H/R_C$ becomes very large. Here,
it is worth to mention the pioneer work on 6LEH spherical hohlraum
fielded at the ISKRA-5 facility with 12 laser beams \cite{Belkov},
in which the ratio  is taken as $R_H/R_C=7$. Obviously, it  costed
double energy as compared to that designed at the golden ratio,
while the corresponding  symmetry was not the best.

In addition to the advantage in high robust symmetry, the golden
octahedral hohlraums  also have superiority on low backscattering
and low plasma filling. As we mentioned above, the spherical
octahedral hohlraums contain only one cone at each LEH, so the
issues of beam overlapping and crossed-beam transfer do not exist.
Thus, the backscattering can be remarkable decreased without any
beam phasing, which therefore  leads to a higher laser absorption
efficiency for a spherical hohlraum than for a cylinder. In
addition,  the volume of a golden octahedral hohlraums  is 125 times
of that capsule volume, more than 2 times of that of the traditional
cylindrical hohlraum which is about 50 to 60 times of the capsule
volume. Thus, the plasma filling inside such an octahedral hohlraum
is obviously lower than inside a cylinder, which further benefit a
low backscattering. Of cause, it needs more laser energy to drive a
larger hohlraum for producing same radiation. Nevertheless, it is
worth to spend some more laser energy to get a robust high symmetry.
As an example, we compare the laser energy required for generating
an ignition radiation pulse of  300 eV   inside a golden octahedral
hohlraum with that inside a traditional cylinder. We consider the
ignition target recently designed for NIF \cite{Kline} and use the
expended plasma-filling model \cite{Dewald, Mcdonald, Lan2012,
Lan2010} to calculate the required laser energy and the plasma
filling inside the hohlraums. According to Ref. \cite{Kline},  the
cylindrical uranium hohlraums with dimensions of 5.75mm in diameter
and 9.4mm in length are used for a DT capsule of $R_C$=1.13 mm. To
have same LEH area, we take the $R_L = $ 1.732 mm for cylinder and
$R_L = $ 1  mm for the golden octahedral hohlraum. Here, we do not
consider the backscattering and assume that the conversion
efficiency from laser to x-ray is $87\%$ for both hohlraums. From
our calculation,
 it needs 1.5 MJ absorbed laser energy by using  the
golden octahedral hohlraum with $n_e = 0.067$, and 1.1 MJ by using
the cylinder with $n_e = 0.094$. Here, $n_e$ is electron density in
unit of the critical density, and the plasma filling criterion is
$n_e=0.1$ \cite{Dewald}. Obviously, it costs more than $30\%$ laser
energy by using a golden octahedral hohlraum, but it is available on
both NIF and LMJ.


\emph{Laser arrangement and constraints}--- We define the hohlraum
pole axis as z axis. Axis x is defined by the centers of two
opposite LEHs on equator, and y is defined by the other two. We name
the LEHs centered on z axis as LEH1 and LEH6, on y axis as LEH2 and
LEH4, and on x axis as LEH3 and LEH5. Each quad through a LEH is
characterized by $\theta _L$ and $\phi _L$.
 There is only one cone in our design, so
all quads coming from the six LEHs have the same $\theta _L$. The
chooses of $\theta _L$ and $\phi _L$ are not only related to the
ratios of $R_H/R_C$, $R_L/R_C$, $R_Q/R_C$ but also interactional.
There are three constraints which govern the quad arrangement.
First, the lasers can not hit to the opposite half sphere in order
to have a short transfer distance inside hohlraum for suppressing
the increase of LPI, which limits the opening angle $\theta _L >
45^\circ$. Second, the laser can not enter the hohlraum at a very
shallow angle in order to avoid absorbing by blowoff from the wall
and making unclearance of the hole. The latter requires $\theta _L <
arcsin((R_H-R_Q)/h)$, here $h=\sqrt {R_H^2 - R_L ^2}$. For model in
Fig.1, it requires $\theta _L < 65^\circ$. Third, a laser beams can
not cross and overlap with other beams.

We use $N_Q$ to denote the quad number per LEH. The quads come in
each LEH coordinately around LEH axis at the azimuthal angles of
$\phi _{L0} + k \times 360^{\circ}/N_Q$ $(k=1, ..., N_Q)$. Here,
$\phi _{L0}$ is azimuthal angle deviated from  x axis in the xy
plane for LEH1 and LEH6, from x axis in the xz plane for LEH2 and
LEH4, and from y axis in the yz plane for LEH3 and LEH5. From the
geometrical symmetry, we have $0^{\circ} < \phi _{L0} <
360^{\circ}/{2N_Q}$. In order to avoid  overlapping between  laser
spots and transferring out of neighbor LEHs, we usually take $\phi
_{L0} $ around $360^{\circ}/{4N_Q}$. In Fig. 1, we take $\phi
_{L0}=11.25^\circ $.

\begin{figure}[t]
\includegraphics[bb=00 250 450 680,width=5 cm]{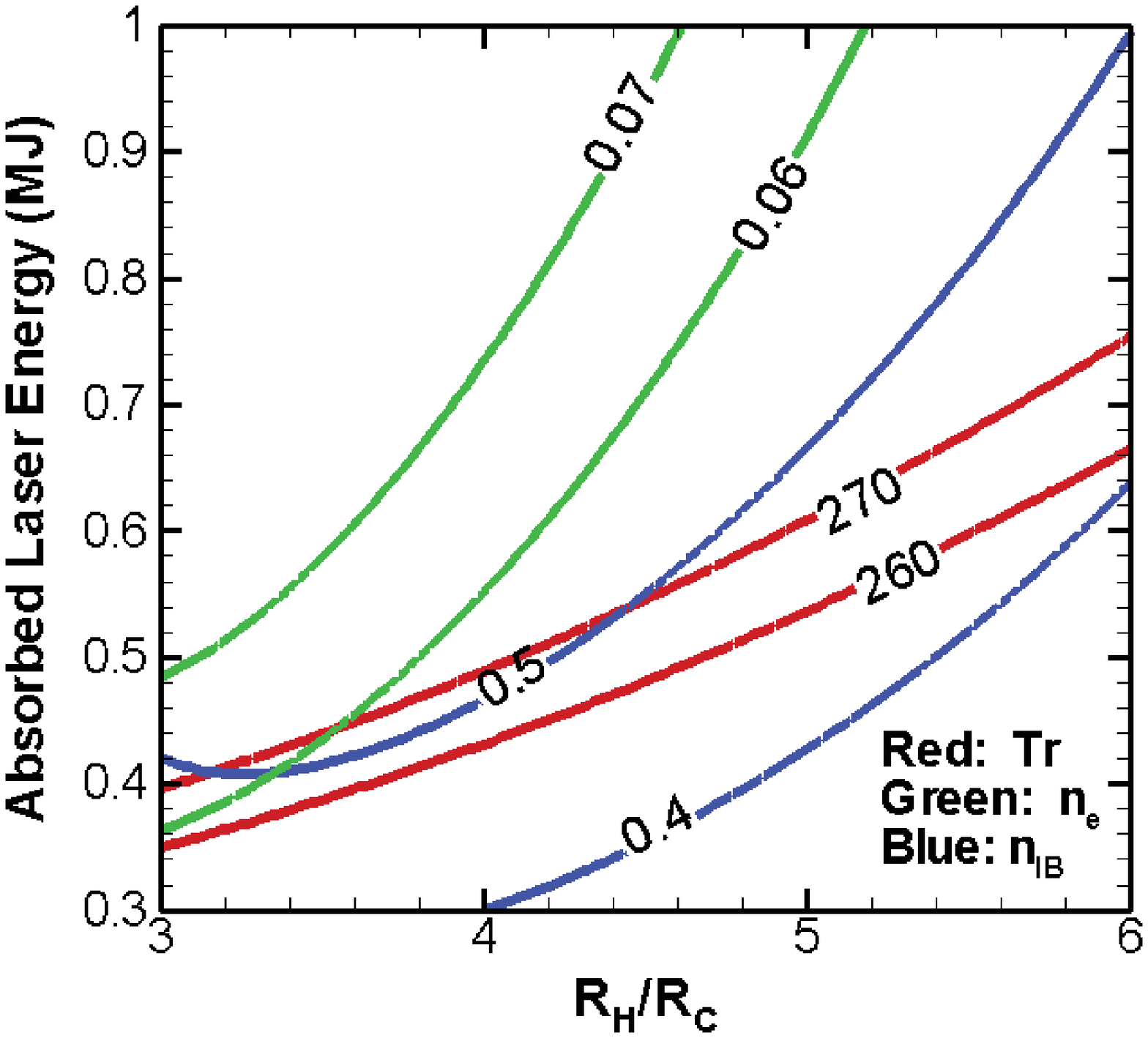}
\caption{(color online) Initial design of laser energy and $R_H
/R_C$to produce the required radiation for the HID model. Red lines
are contours of $T_r=$260 eV and 270 eV and green lines are contours
of $n_e=$ 0.05 and 0.07, and blue lines are contours of $n_{IB}$ 0.4
and 0.5. } \label{Fig5}
\end{figure}

\emph{Application}---The proposed octahedral hohlraum is  flexible
and can be applicable to diverse inertial fusion drive approaches
such as  indirect and hybrid indirect-direct drives. As an
application, we design a golden octahedral hohlraum for the hybrid
drive by using the expended plasma-filling model  and  view factor
model. In the HID model\cite{Fan}, the fuel capsule is first
compressed by indirect-drive x rays, and then by both x rays and
direct drive lasers. According to the HID model, a four-step
radiation pulse with the fourth step of 260 to 270 eV and 1.7 ns  is
required for a capsule with radius $R_C$ = 850 $\mu$m. Shown in Fig.
5 is the initial design of laser energy and $R_H$ by using the
extended plasma-filling model. Here, $R_L$ is taken as 1 mm. The two
semi-empirical criterions used here are: $n_e \leq 0.1$  and $n_{IB}
\equiv (l/\sqrt{2})/(\lambda _{IB}) \leq 1 $. The latter is related
to the inverse bremsstrahlung absorption length $\lambda _{IB}$
\cite{Dawson}. Here,   $l$ is the transfer distance of laser beam
inside a spherical hohlraum.  As shown, it needs absorbed laser
energy of 0.5 to 0.6 MJ to produce a 260 to 270 eV radiation pulse
inside a golden octahedral hohlraum  with $n_e$ and $n_{IB}$ well
meeting the criterions. The dependence of the capsule asymmetry on
$R_{H}/R_{C}$,  $R_L$, $R_Q$, $\theta _L$ and the relative flux of
laser spot to hohlraum wall is important for choosing the optimum
design of hohlraum. Using the view factor model, we study the
variations of $|\Delta F /\langle F \rangle|$ as these quantities
for the HID model. The results indicate that $|\Delta F /\langle F
\rangle|$ is smaller than $0.2\%$ at the golden ratio in the ranges
of $R_{H}/R_{C}$, $R_L$, $R_Q$ and $\theta _L$ concerned in our
model.


In summary, we have investigated the spherical hohlraum with
octahedral 6LEHs at a golden hohlraum-to-capsule radius ratio, which
has very high and robust symmetry on capsule, with significantly
lower plasma filling and lower backscattering  as compared to the
cylindrical counterpart. Inside a golden octahedral hohlraum, it is
$Y_{8m}$  which dominates the asymmetry. It needs
 about $30\%$ more laser energy to drive such a golden
octahedral hohlraum than to drive a cylinder for producing same
radiation, but it is worth to exchange such available laser energy
for  a robust high symmetry. The above novel spherical hohlraum
design has important implications for laser inertial fusion and is
expected to be conducted on SG laser facilities in near future.


\begin{acknowledgements}
The authors wish to acknowledge the beneficial help from Dr. Yiqing
Zhao, Dr. Zhengfeng Fan, Prof. Xiaomin Zhang, Prof. Juergen
Meyer-ter-Vehn, Prof. Grant Logan and Prof. Min Yu. This work is
supported by the National Fundamental Research Program of China
(Contact No. 2013CBA01502, 2013CB83410, and 2013A0102002).

\end{acknowledgements}


\begin{thebibliography}{22}





\bibitem{MTV} S. Atzeni, J. Meyer-ter-Vehn, The Physics of Inertial Fusion (Oxford Science, Oxford, 2004).

\bibitem{lindl2004} J. D. Lindl, Phys. Plasmas {\bf 2}, 3933 (1995).

\bibitem{Haan}S. W. Haan et al., Phys. Plasmas {\bf 18}, 051001 (2011).

\bibitem{Fan} Zhengfeng Fan et al.,
arXiv:1303.1252[physics.plasm-ph]; X. T. He, the plenary
presentation at IFSA 8, September 9-13, 2013, Nara, Japan.

\bibitem{Caruso} A. Caruso and C. Strangio, Japanese Journal of Applied Physics
{\bf 30}, 1095 (1991).

\bibitem{Amendt}  P. Amendt et al., Phys. Plasmas {\bf 14}, 056312(2007).

\bibitem{Vandenboomgaerde} M. Vandenboomgaerde et al., Phys. Rev. Lett. {\bf 99}, 065004 (2007).

\bibitem{Casner} A. Casner et al., Phys. Plasmas {\bf 16}, 092701 (2009).

\bibitem{Robey} H.F. Robey et al., Phys. Plasmas {\bf 17}, 056313 (2010).

\bibitem{Phillippe} F. Philippe et al., Phys. Rev. Lett. {\bf 104}, 035004 (2010).

\bibitem{Lan2012} K. Lan, et al., Laser and Particle Beams {\bf 30}, 175 (2012).



\bibitem{Callahan} D.A. Callahan et al., Phys. Plasmas {\bf 19}, 056305 (2012).

\bibitem{Kline}  J. L. Kline et al., Phys. Plasmas {\bf 20}, 056314 (2013).



\bibitem{Belkov} S. A. Bel'kov et al., JEPT Lett. {\bf 67}, 171 (1998).


\bibitem{Wallace} J. M. Wallace, et al., Phys. Rev. Lett. 82, 3807 (1999).

\bibitem{Phillion} D. W. Phillion and S. M. Pollaine, Phys. Plasmas {\bf 1}, 2963
(1994).






\bibitem{Dewald} E.L. Dewald et al., Phys. Rev. Lett. {\bf 95}, 215004 (2005).

\bibitem{Mcdonald} J.W. Mcdonald et al., Phys. Plasmas {\bf 13}, 032703(2006).

\bibitem{Lan2010}K. Lan et al., Laser and Particle Beams {\bf 28}, 421(2010).



\bibitem{Dawson}J. Dawson, P. Kaw, and B. Green, Phys. Fluids 12, 875(1969).




\end{thebibliography}
\end{document}